\documentclass[12pt]{iopart}

\usepackage{lineno,hyperref}
\expandafter\let\csname equation*\endcsname\relax
\expandafter\let\csname endequation*\endcsname\relax
\usepackage{amsmath}
\usepackage{epstopdf}
\usepackage{graphicx}
\usepackage{color} 
\usepackage{tabularx}
\usepackage{subfigure}
\usepackage{latexsym}
\usepackage{longtable}
\usepackage{psfrag}
\usepackage{epsfig}
\usepackage{graphicx,subfigure}
\usepackage{array}
\usepackage{tabularx}
\usepackage{hyperref}

\begin{document}

\title[QFI protection of $N$-qubit GHZ state from decoherence]{Quantum fisher information protection of $N$-qubit Greenberger-Horne-Zeilinger state from decoherence}

\author{Sajede Harraz\textsuperscript{1}, Shuang Cong\textsuperscript{1}, Jiaoyang Zhang\textsuperscript{1} and Juan J. Nieto\textsuperscript{2}}

\address{$^1$ Department of Automation, University of Science and Technology of China, Hefei 230027, China}

\address{$^2$ Instituto de Matematicas, CITMAga, Universidade de Santiago de Compostela, 15782 Santiago de Compostela, Spain}

\ead{sajede@ustc.edu.cn}

\vspace{10pt}

\begin{abstract}
In this paper we study the protection of \textit{N}-qubit Greenberger-Horne-Zeilinger (GHZ) state and generalized \textit{N}-qubit GHZ states in amplitude damping channel by means of quantum weak measurement and flip operations. We derive the explicit formulas of the performances of the protection scheme: average fidelity, average probability and the average quantum fisher information (QFI). Moreover, the analytical results for maximizing the average fidelity and probability are obtained. We show that our scheme can effectively protect the average QFI of phase for GHZ states and generalized GHZ states. The proposed scheme has the merit of protecting GHZ state and the QFI of phase against heavy amplitude damping noise. Further we show that for some generalize GHZ state, the proposed scheme can protect the state with probability one and fidelity more than 99\%.
\end{abstract}

%
\vspace{2pc}

\section{Introduction}

One of the basic tasks in information theory is parameter estimation of probability distributions, which has been generalized to quantum systems \cite{lab4,lab5}. The Fisher information as a central role in estimation theory measures the amount of information of a parameter that we can extract from a probability distribution. By larger fisher information one can estimate the parameter more accurately. Since the characteristic of quantum mechanics is probabilistic, fisher information can be extended to quantum regime. Quantum fisher information (QFI) is defined in Cramer-Rao bound $\delta \phi \ge {1\mathord{\left/ {\vphantom {1 \sqrt{vF} }} \right. \kern-\nulldelimiterspace} \sqrt{vF} } $, where $\phi $ is the parameter to be estimated, $v$ the measurement times and F is the QFI \cite{lab6}. Hence, obtaining large QFI is a crucial task in quantum control. However, the inevitable interaction with environment in real quantum systems, which cause decoherence, needs to be taken into account \cite{lab7}. One of the important noise sources is amplitude damping which occurs in many quantum systems \cite{lab8}, such as a photon qubit in a leaky cavity, an atomic qubit subjected to spontaneous decay, or a super-conduction qubit with zero-temperature energy relaxation. Various strategies are studied to overcome the effects of decoherence, such as: quantum error correction \cite{lab9,lab10,lab11,lab12,lab13,lab14}, decoherence-free subspaces \cite{lab15,lab16}, quantum feedback control (QFBC) \cite{lab17,lab18,lab19,lab20,lab21,lab22} and quantum feed-forward control (QFFC) \cite{lab23,lab24,lab25}.  To protect the state from environmental noise, a quantum feedback control was proposed by using weak measurements and rotation operation after the noise channel in \cite{lab18}. This scheme is experimentally implemented in \cite{lab19} and is considered for protection of arbitrary initial states against different types of noise sources \cite{lab26,lab27}. Quantum feedback control applies control operations after the noise channel, while quantum feed-forward control is about control operations both before and after noise channel. The aim to apply operations before noise is to make the state immune to the effects of the noise channel. Hence, one needs to know the characteristics of the noise. In \cite{lab23} a quantum feed-forward control is proposed to protect the state from amplitude damping, and was extended to protect two-qubit systems from amplitude damping channel (ADC) \cite{lab28}. This scheme is also considered for discrimination of two non-orthogonal states against noise \cite{lab25}. 

In quantum systems, any measurement comes with a price which is disturbance of the system \cite{lab29,lab30}. However, by quantum weak measurement one can find a trade-off between information gain and disturbing it through measurement. The basic idea of the weak measurement is that the interaction (or disturbance) between the measuring apparatus and the observed system or particle is so weak, that the wave function does not collapse but continues unchanged \cite{lab31}. In both QFBC and QFFC schemes, the weak measurement with an optimum measurement strength is used to find the optimal protection from noise. 

In this paper we study the protection of the average QFI and the state of \textit{N}-qubit Greenberger--Horne--Zeilinger (GHZ) state from ADC by means of weak measurement and offsetting operations. The control procedure consists of pre and post noise channel operations. Hence, we call the proposed scheme as quantum feed-forward control and its reversal (QFFCR) hereafter. Before noise channel, a weak measurement applies to gain information about the state of the system. Then according to the result of the measurement, we apply an offsetting operation to make the state less vulnerable to the effect of ADC. After the noise channel, the aim is to retrieve the state, so the same offsetting operation as the one before the noise channel is applied. Finally, to bring the state as close as possible to its initial state we apply a correction rotation at the final step of the protection control. Although the final states of the scheme for \textit{N}-qubit is different (there are $2^N$ kind results of the final state which come from the $2^N$ kind measurement results of the weak measurement), they have the same structure, so we can calculate the average QFI, average fidelity and average success probability analytically. We derive the final expressions for three performances of the control: average total QFI, fidelity and probability of \textit{N}-qubit generalized GHZ states; and study the optimum parameters to gain maximum of each performances of the proposed control scheme. Our scheme can protect not only GHZ states but also some generalized GHZ states. The simulation results show that by using optimum parameters, the proposed QFFCR scheme can protect some generalized GHZ states with probability 1 and fidelity close to 1 even for intense decaying rates. 

For comparison, we discuss another scheme that uses weak-measurement-based pre- and post-flips (WMPPF) \cite{lab33} and demonstrate by the simulation results that our scheme has much better performance. In WMPPF scheme the state is measured by WM operators before the noise channel, and the flips are applied according to the result of the WM to make the state almost immune to the effects of the noise. Then, after the noise channel the reversed flips are applied to recover the state. The WMPPF has the success probability equal to 1 for all decaying rates and measurement strengths while the QFFCR is a probabilistic scheme.

This paper is organized as follows: In Sec. 2, we define the QFFCR scheme for one-qubit protection. Sec. 3 is the protection scheme performance derivations of $N$-qubit generalized GHZ state, including average total probability, fidelity and QFI. In Sec. 4, we do the numerical simulations and study the behavior of QFFCR performances for protection of $N$-qubit GHZ state in ADC. Finally, in Sec. 5 we give the conclusion of our results. 

\section{Noise protection control scheme for one-qubit}

The protection scheme is divided into two parts, the operations before the noise channel, which are the weak measurement and offsetting operations; and the operations after the noise channel, which are the offsetting and rotation operations, respectively. The aim of before noise operations is to make the state almost immune to the effect of noise channel and by after noise operations, we try to retrieve the state and bring it back as close as possible to its initial state.

\begin{figure}[h]
    \centering
    \includegraphics[width=1\textwidth]{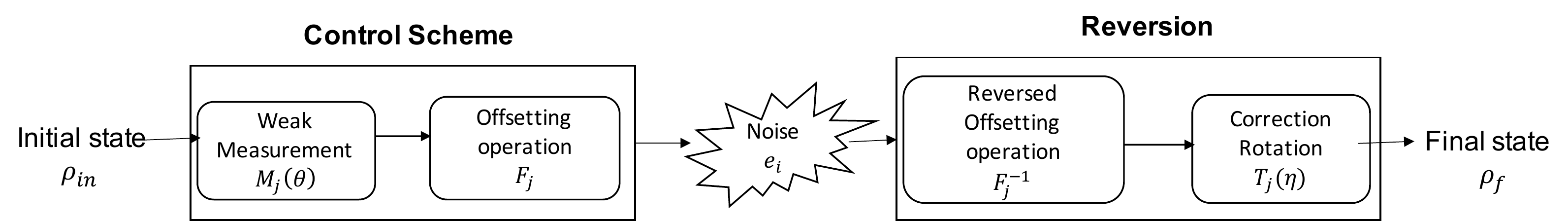}
  
\caption{Schematic diagram of QFFCR.}\label{fig1}
\end{figure}

The whole process of state protection consists of 5 steps, which is shown in Fig. \ref{fig1} and  is explained in the following \cite{lab32}. 

\textbf{Step 1:} In the first step, we apply the weak measurement as $\Pi _{0} =M_{0}^{\dag } M_{0} $ and $\Pi _{1} =M_{1}^{\dag } M_{1} $ to gain some information about the system. The weak measurement operators are along \textit{z}-axis of the Bloch sphere as:

\begin{equation} \label{eq1} 
\begin{array}{l} {M_{0} =\cos (\theta /2){\left| 0 \right\rangle} {\left\langle 0 \right|} +\sin (\theta /2){\left| 1 \right\rangle} {\left\langle 1 \right|} } \\ {M_{1} =\sin (\theta /2){\left| 0 \right\rangle} {\left\langle 0 \right|} +\cos (\theta /2){\left| 1 \right\rangle} {\left\langle 1 \right|} } \end{array} 
\end{equation} 
where $\theta \in \left[0,\frac{\pi }{2} \right]$ and the strength of the measurement is defined by the angle $\theta $. When $\theta ={\raise0.7ex\hbox{$ \pi  $}\!\mathord{\left/ {\vphantom {\pi  2}} \right. \kern-\nulldelimiterspace}\!\lower0.7ex\hbox{$ 2 $}} $, the strength of the measurement is zero, and we call it as: ``no measurement'' (NM) scheme. When $\theta =0$, the strength of the measurement is maximum and is called as ``projective measurement'' (PM), in a way that the state of the system projected onto the state ${\left| 0 \right\rangle} $ (or ${\left| 1 \right\rangle} $) when the measurement result is 0 (or 1). When $0<\theta <{\raise0.7ex\hbox{$ \pi  $}\!\mathord{\left/ {\vphantom {\pi  2}} \right. \kern-\nulldelimiterspace}\!\lower0.7ex\hbox{$ 2 $}} $ , we call it as ``weak measurement'' (WM) scheme. Hence, we can adjust the strength of the measurement for the trade-off between the information gain and disturbance of the system because of measurement.

\textbf{Step 2:} According to the result of the measurement, the offsetting operations are applied to make the state immune to the effect of noise before ADC. 
\begin{equation} \label{eq2} 
F_{0}^{} =I^{} =\left(\begin{array}{cc} {1} & {0} \\ {0} & {1} \end{array}\right),\, \, F_{1}^{} =\sigma _{x}^{} =\left(\begin{array}{cc} {0} & {1} \\ {1} & {0} \end{array}\right) 
\end{equation} 
  where $F_{0}^{} \, (\, F_{1}^{} )$ corresponds to the result of the measurement operator $M_{0} \, (M_{1} )$ respectively. When the result corresponds to measurement operator $M_{0} $ is acquired, the state is close to the ground state, so noise will not have much effects on the state. Therefore, in offsetting step we do not need to make changes in the state, and $F_{0} $ is chosen as identity operator. However, when the result corresponds to measurement operator $M_{1} $ is acquired, by applying the operator $F_{1} $,  the state of the system becomes closer to the ground state and less vulnerable to the effects of noise.
  
  \textbf{    Step 3: }In this step, the state of the system is ready to enter the ADC. For each qubit the amplitude damping is represented by Kraus operators as:
\begin{equation} \label{eq3} 
e_{0} =\left(\begin{array}{cc} {1} & {0} \\ {0} & {\sqrt{1-r} } \end{array}\right),\, e_{1} =\left(\begin{array}{cc} {0} & {\sqrt{r} } \\ {0} & {0} \end{array}\right) 
\end{equation} 
where $r\in \left[0,1\right]$ is the possibility of decaying of the excited state with $r=e^{-\Gamma t} $, $\Gamma $ is the energy relaxation rate and $t$ is the evolving time \cite{lab34}.

 \textbf{   Step 4:} After the ADC, the same offsetting operations as given in Eq. \eqref{eq2} are applied. 

\textbf{   Step 5:} At the final step, to correct the state and bring it back to its initial state, the correction rotation is applied as:
\begin{equation} \label{eq4)} 
T_{0}^{} =\left(\begin{array}{cc} {e^{i\frac{\eta }{2} } } & {0} \\ {0} & {e^{-i\frac{\eta }{2} } } \end{array}\right)\, \, ,\, \, \, T_{1}^{} =\left(\begin{array}{cc} {e^{-i\frac{\eta }{2} } } & {0} \\ {0} & {e^{i\frac{\eta }{2} } } \end{array}\right)\,  
\end{equation} 
where $T_{0}^{} \, (\, T_{1}^{} )$ corresponds to the result of measurement operator $M_{0} \, (M_{1} )$, respectively.

\section{ Protection scheme Performance Derivations of N-qubit generalized GHZ state}

Here we give the average QFI, average fidelity and average probability of proposed QFFCR for general \textit{N}-qubit GHZ state. The initial state in this paper is the \textit{N-}qubit generalized GHZ state as 
\begin{equation} \label{eq5} 
{\left| \psi _{in}  \right\rangle} =\alpha {\left| 0 \right\rangle} ^{\otimes N} +\beta {\left| 1 \right\rangle} ^{\otimes N}  
\end{equation} 
 where $\alpha =\cos \left({\gamma \mathord{\left/ {\vphantom {\gamma  2}} \right. \kern-\nulldelimiterspace} 2} \right),\, \beta =e^{i\phi _{0} } \sin \left({\gamma \mathord{\left/ {\vphantom {\gamma  2}} \right. \kern-\nulldelimiterspace} 2} \right),\, 0<\gamma <1$, and $\phi _{0} $ is the initial phase. By setting $\gamma ={\pi \mathord{\left/ {\vphantom {\pi  2,\, \, \, \phi _{0} =0}} \right. \kern-\nulldelimiterspace} 2,\, \, \, \phi _{0} =0} $ we gain the GHZ state. Once more, we assume that all the qubits have the same control parameters, same weak measurement strength, same damping probability and same rotation angle. 

The normalized final state of QFFCR is attainable, in which there are $2^{N} $different final states according to $2^{N} $ different kind of measurement results. Fortunately, all different results of final states have the same structure as
\begin{equation} \label{eq6} 
\rho _{f} =\frac{1}{P} \tilde{\rho }=\frac{1}{P} \left(\begin{array}{ccc} {A} & {0} & {D} \\ {0} & {E} & {0} \\ {C} & {0} & {B} \end{array}\right) 
\end{equation} 

where $\tilde{\rho }$ is the unnormalized final state and $P$ is the average total probability of appearing $\rho _{f} $. 

The detailed calculation of the elements of \textit{A, B, C, D }and \textit{E} matrix of \textit{N}-qubit GHZ state in  Eq. \eqref{eq6} at each step in QFFCR is given in \textbf{Appendix}. Here we calculate the average total probability, fidelity and QFI. 

\subsection{ Average total probability}

For the normalized final state Eq. \eqref{eq6} the total probability is calculated as
\begin{equation} \label{eq7)} 
P=Tr\left(\tilde{\rho }\right)=A+B+Tr(E) 
\end{equation} 
Hence, the average total probability is the sum of the probabilities of three cases, $k=0$, $k=N$ and $k=1$ to $N-1$. 
\begin{equation} \label{8)} 
\begin{aligned}
 P_{0} &=A_{0} +B_{0} +Tr(E_{0} )=\\&\alpha ^{2} \sin ^{2N} \left(\frac{\theta }{2} \right)\left(re^{Ni\eta } +\left(1-r\right)e^{-Ni\eta } \right)^{N} +\beta ^{2} \cos ^{2N} \left(\frac{\theta }{2} \right)e^{Ni\eta }  \\ P_{N} &=A_{N} +B_{N} +Tr(E_{N} )=\\&\alpha ^{2} \cos ^{2N} \left(\frac{\theta }{2} \right)e^{Ni\eta } +\beta ^{2} \sin ^{2N} \left(\frac{\theta }{2} \right)\left(re^{i\eta } +\left(1-r\right)e^{-i\eta } \right)^{N}  \\ P_{S_{k} } &=A_{S_{k} } +B_{S_{k} } +Tr(E_{S_{k} } )=\\&\alpha ^{2} \sin ^{2\left(N-k\right)} \left(\frac{\theta }{2} \right)\cos ^{2k} \left(\frac{\theta }{2} \right)e^{ki\eta } \left(re^{i\eta } +\left(1-r\right)e^{-i\eta } \right)^{N-k} + \\ &\beta ^{2} \cos ^{2\left(N-k\right)} \left(\frac{\theta }{2} \right)e^{\left(N-k\right)i\eta } \sin ^{2k} \left(\frac{\theta }{2} \right)\left(re^{i\eta } +\left(1-r\right)e^{-i\eta } \right)^{N-k} 
\end{aligned}
\end{equation} 

When $k=1$ to $N-1$, there are ${\rm C}_{N}^{k} $  kind cases, so the average total probability becomes
\begin{equation} \label{eq9} 
\begin{aligned}
 P_{total} =P_{0} +P_{N} &+\sum _{k=1}^{N-1}\sum _{lk=1}^{{\rm C}_{N}^{k} }P_{lk} =\left(re^{i\eta } +\left(1-r\right)e^{-i\eta } \right)^{N} \sin ^{2N} \left(\frac{\theta }{2} \right)+\\&\cos ^{2N} \left(\frac{\theta }{2} \right)e^{iN\eta } +\sum _{k=1}^{N-1}\frac{N!}{k!(N-k)!}  P_{lk} =\\&\left(\left(re^{i\eta } +\left(1-r\right)e^{-i\eta } \right)\sin ^{2} \left(\frac{\theta }{2} \right)+e^{i\eta } \cos ^{2} \left(\frac{\theta }{2} \right)\right)^{N}  
\end{aligned}
\end{equation} 

\subsection{ Average total fidelity}

The fidelity between initial state ${\left| \psi _{in}  \right\rangle} $ Eq. \eqref{eq5} and final state $\rho _{f} $  Eq. \eqref{eq6} is given as
\begin{equation} \label{10)} 
Fid=\left\langle \psi _{in} \right. \left|\rho _{f} \right|\left. \psi _{in} \right\rangle =\frac{1}{P} \left(\left|\alpha \right|^{2} A+\alpha \beta ^{*} C+\alpha ^{*} \beta D+\left|\beta \right|^{2} B\right) 
\end{equation} 
where $P$ is the average total probability. The fidelities in three cases of  $k=0$, $k=N$ and $k=1$ to $N-1$ are 
\begin{equation} \label{eq11)} 
\begin{array}{l} {Fid_{0} =\frac{1}{p_{0} } \left(\left|\alpha \right|^{2} A_{0} +\alpha \beta ^{*} C_{0} +\alpha ^{*} \beta D_{0} +\left|\beta \right|^{2} B_{0} \right);} \\ {Fid_{N} =\frac{1}{p_{0} } \left(\left|\alpha \right|^{2} A_{N} +\alpha \beta ^{*} C_{N} +\alpha ^{*} \beta D_{N} +\left|\beta \right|^{2} B_{N} \right);} \\ {Fid_{S_{k} } =\frac{1}{p_{0} } \left(\left|\alpha \right|^{2} A_{S_{k} } +\alpha \beta ^{*} C_{S_{k} } +\alpha ^{*} \beta D_{S_{k} } +\left|\beta \right|^{2} B_{S_{k} } \right);} \end{array} 
\end{equation} 
Therefore, the average total fidelity is the sum of the fidelities of three cases, $k=0$, $k=N$ and $k=1$ to $N-1$ as:
\begin{equation} \label{eq12} 
\begin{array}{l} {Fid_{total} =\frac{1}{p_{total} } \left[P_{0} Fid_{0} +P_{N} Fid_{N} +\sum _{k=1}^{N-1}\sum _{S_{k} =1}^{{\rm C}_{N}^{k} }P_{S_{k} } Fid_{S_{k} }   \right]} \\ {=\frac{1}{p_{total} } \left[\begin{array}{l} {\left(\left|\alpha \right|^{2} A_{0} +\alpha \beta ^{*} C_{0} +\alpha ^{*} \beta D_{0} +\left|\beta \right|^{2} B_{0} \right)+
}\\{\left(\left|\alpha \right|^{2} A_{N} +\alpha \beta ^{*} C_{N} +\alpha ^{*} \beta D_{N} +\left|\beta \right|^{2} B_{N} \right)+} \\ {\sum _{k=1}^{N-1}\sum _{S_{k} =1}^{{\rm C}_{N}^{k} }\left(\left|\alpha \right|^{2} A_{S_{k} } +\alpha \beta ^{*} C_{S_{k} } +\alpha ^{*} \beta D_{S_{k} } +\left|\beta \right|^{2} B_{S_{k} } \right)  } \end{array}\right]} \\ {=\frac{1}{p_{total} } \left[\begin{array}{l} {\left(\left|\alpha \right|^{4} +\left|\beta \right|^{4} \right)\left[\cos ^{2} \left(\frac{\theta }{2} \right)e^{i\eta } +\sin ^{2} \left(\frac{\theta }{2} \right)e^{-i\eta } \left(1-r\right)\right]^{N} +}\\{2\left|\alpha \beta \right|^{2} r^{N} e^{iN\eta } \sin ^{2N} \left(\frac{\theta }{2} \right)} \\ {+2^{N+1} \left|\alpha \beta \right|^{2} \left(1-r\right)^{{\raise0.7ex\hbox{$ N $}\!\mathord{\left/ {\vphantom {N 2}} \right. \kern-\nulldelimiterspace}\!\lower0.7ex\hbox{$ 2 $}} } \frac{\sin ^{N} (\theta )}{2^{N} } } \end{array}\right]} \end{array} 
\end{equation} 

As the final expression of average total fidelity Eq. \eqref{eq12} shows, there is a trade-off between average total fidelity and probability, so by increasing one, the other will decrease. Hence, one needs to find the best balance between fidelity and probability. 

\subsection{ Average total QFI}

In classical definition, fisher information is measuring the amount of information that an observable random variable carries about an unknown parameter $\sigma $. The quantum fisher information is formally generalized from classical fisher information definition as \cite{lab35,lab36}:
\begin{equation} \label{eq13} 
F_{\sigma } =\sum _{l'}\frac{\left(\partial _{\sigma } \lambda _{l'} \right)^{2} }{\lambda _{l'} } +\sum _{l\ne m}\frac{2\left(\lambda _{l} -\lambda _{m} \right)^{2} }{\lambda _{l} +\lambda _{m} } \left|\left\langle \varphi _{l} \right. \left|\partial _{\sigma } \right|\left. \varphi _{m} \right\rangle \right|^{2}    
\end{equation} 
where ${\partial }_{\sigma }\equiv \frac{\partial }{\partial \sigma }$  and ${\left| \varphi _{m}  \right\rangle} $ , ${\left| \varphi _{l}  \right\rangle} $ are the eigenvectors and $\lambda $ the eigenvalues of the state. The first summation is equal to the classical fisher information which is called the classical term and the second summation is called the quantum term. 

In this paper, we consider the parameter $\phi _{0}$ phase of the initial state in Eq. \eqref{eq5}, as the unknown parameter to be measured and estimated.

As it has been proved in \cite{lab33}, to calculate the QFI of phase $\phi _{0} $ for density matrix with the structure of Eq. \eqref{eq6}, Eq. \eqref{eq13} can be simplified as:
\begin{equation} \label{14)} 
QFI=\frac{1}{P} \frac{4\left|C\right|^{2} }{A+B} N^{2}  
\end{equation} 
where $A,\ B\ \mathrm{and}\ C$ are the density matrix elements as given in Eq. \eqref{eq6} and $N$ the number of qubits.

Hence the QFI in three cases of $k=0$, $k=N$ and $k=1$ to $N-1$ are 
\begin{equation} \label{eq15} 
QFI_{0} =\frac{1}{P_{0} } \frac{4\left|C_{0} \right|^{2} N^{2} }{A_{0} +B_{0} } ,\, \, QFI_{N} =\frac{1}{P_{N} } \frac{4\left|C_{N} \right|^{2} N^{2} }{A_{N} +B_{N} } ,\, \, QFI_{S_{k} } =\frac{1}{P_{S_{k} } } \frac{4\left|C_{S_{k} } \right|^{2} N^{2} }{A_{S_{k} } +B_{S_{k} } }  
\end{equation} 

Therefore, the average total QFI of phase of our QFFCR scheme is 

\begin{equation} \label{eq16} 
\begin{array}{l} {QFI_{total} =P_{0} QFI_{0} +P_{N} QFI_{N} +\sum _{k=1}^{N-1}\sum _{S_{k} =1}^{{\rm C}_{N}^{k} }P_{S_{k} } QFI_{S_{k} }   } \\ {=\frac{4\left|\alpha \beta \right|^{2} \left(1-r\right)^{N} \frac{\sin ^{2N} \left(\theta \right)}{2^{2N} } N^{2} }{\left|\alpha \right|^{2} \sin ^{2N} \left(\frac{\theta }{2} \right)\left(r\left(1-r\right)\right)^{N} +\left|\beta \right|^{2} \cos ^{2N} \left(\frac{\theta }{2} \right)e^{Ni\eta } } } \\ {+\frac{4\left|\alpha \beta \right|^{2} \left(1-r\right)^{N} \frac{\sin ^{2N} \left(\theta \right)}{2^{2N} } N^{2} }{\left|\alpha \right|^{2} \cos ^{2N} \left(\frac{\theta }{2} \right)e^{Ni\eta } +\left|\beta \right|^{2} \sin ^{2N} \left(\frac{\theta }{2} \right)\left(r\left(1-r\right)\right)^{N} } +\sum _{k=1}^{N-1}\frac{N!}{k!(N-k)!}  } \\ {\left(\frac{4\left|\alpha \beta \right|^{2} \left(1-r\right)^{N} \frac{\sin ^{2N} \left(\theta \right)}{2^{2N} } N^{2} }{\left|\alpha \right|^{2} \cos ^{2k} \left(\frac{\theta }{2} \right)\sin ^{2\left(N-k\right)} \left(\frac{\theta }{2} \right)\left(1-r\right)^{N-k} e^{\left(-N+2k\right)i\eta } +\left|\beta \right|^{2} \sin ^{2k} \left(\frac{\theta }{2} \right)\left(1-r\right)^{k} e^{\left(N-2k\right)i\eta } \cos ^{2\left(N-k\right)} \left(\frac{\theta }{2} \right)} \right)} \end{array} 
\end{equation} 
where $\theta \in \left[0,\frac{\pi }{2} \right]$ is the measurement strength, $r\in \left[0,1\right]$ is the decay rate, $N$ is the number of qubits and $k=0$ to $N$.

\section{ QFFCR performances for protection of N-qubit GHZ state in ADC }

In this section, the experimental simulation results and analysis of proposed QFFCR protection scheme for \textit{N}-qubit GHZ state against amplitude damping are given. 

As we mentioned before, according to Eq. \eqref{eq12} there is a trade-off between average total probability and average total fidelity. By maximizing one, we get the lowest amount of the other. Hence, we study maximizing each one separately.

According to Eq. \eqref{eq9}, Eq. \eqref{eq12} and Eq. \eqref{eq16}, the amount $\left|\beta \right|$ is appeared in the final derived equations of performances; Hence any value for $\phi _{0} $ will get the same result for the total average probability, fidelity and QFI. Therefore, we do not consider the value of $\phi _{0} $ in numerical simulations. 

First, we do the simulation experiments to find the behavior of average total QFI. The initial state is given in Eq. \eqref{eq5}, which is GHZ state when $\gamma ={\pi }/{2}\ $ and is generalized GHZ state (GGHZ) state when$\ \gamma \neq {\pi }/{2}$. To show the effectiveness of our scheme, we give the performance comparison with WMPPF scheme proposed in \cite{lab33}. 

\subsection{ Maximum performances}

\subsubsection{ Maximum QFI}

\begin{figure*}[ht!]
    \centering
    \begin{minipage}[t]{0.32\textwidth}
        \centering{\includegraphics[width=\textwidth]{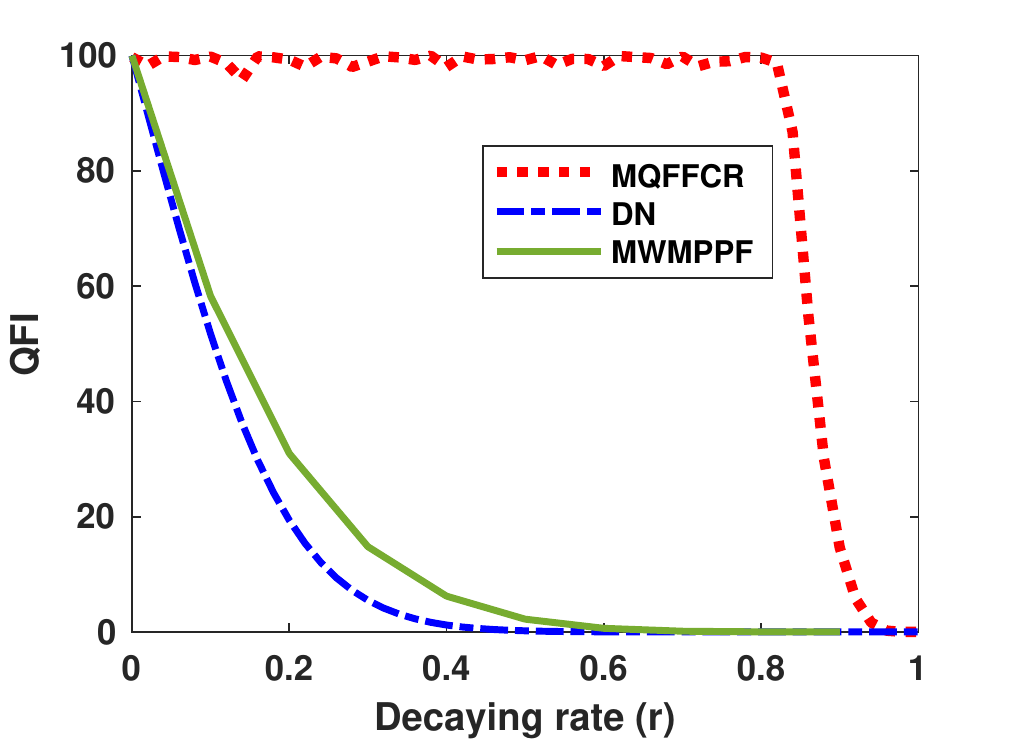}}
       \centering{(a)}
    \end{minipage}
    \hfill
    \begin{minipage}[t]{0.32\textwidth}
        \centering{\includegraphics[width=\textwidth]{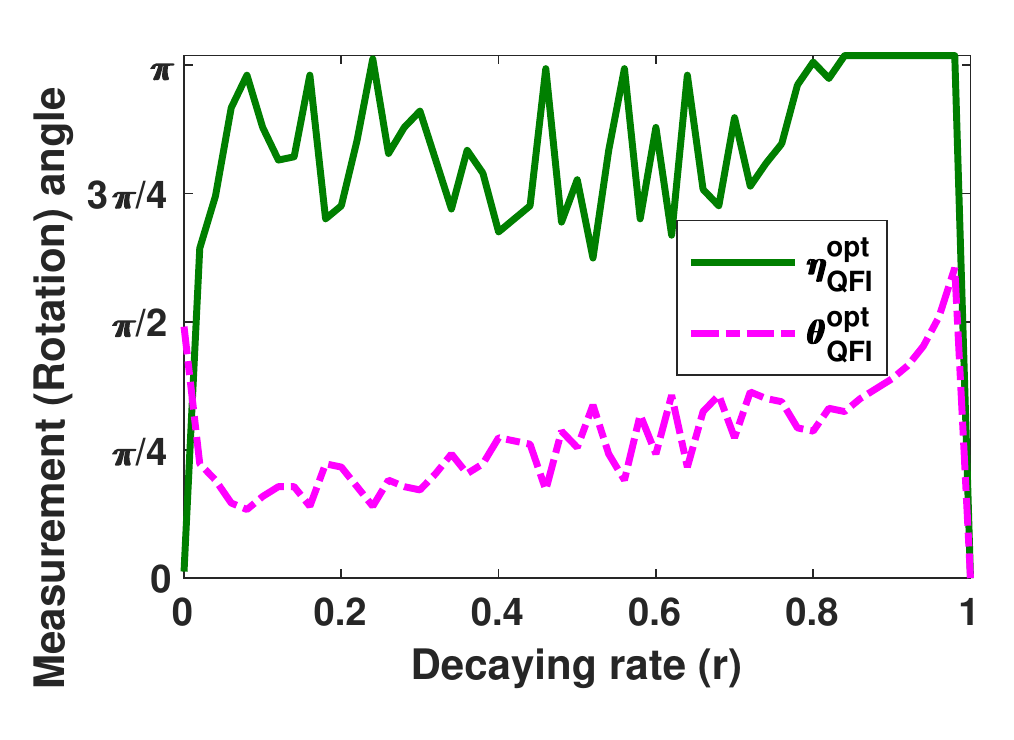}}
      	\centering{(b)}
    \end{minipage}
    \begin{minipage}[t]{0.32\textwidth}
        \centering{\includegraphics[width=\textwidth]{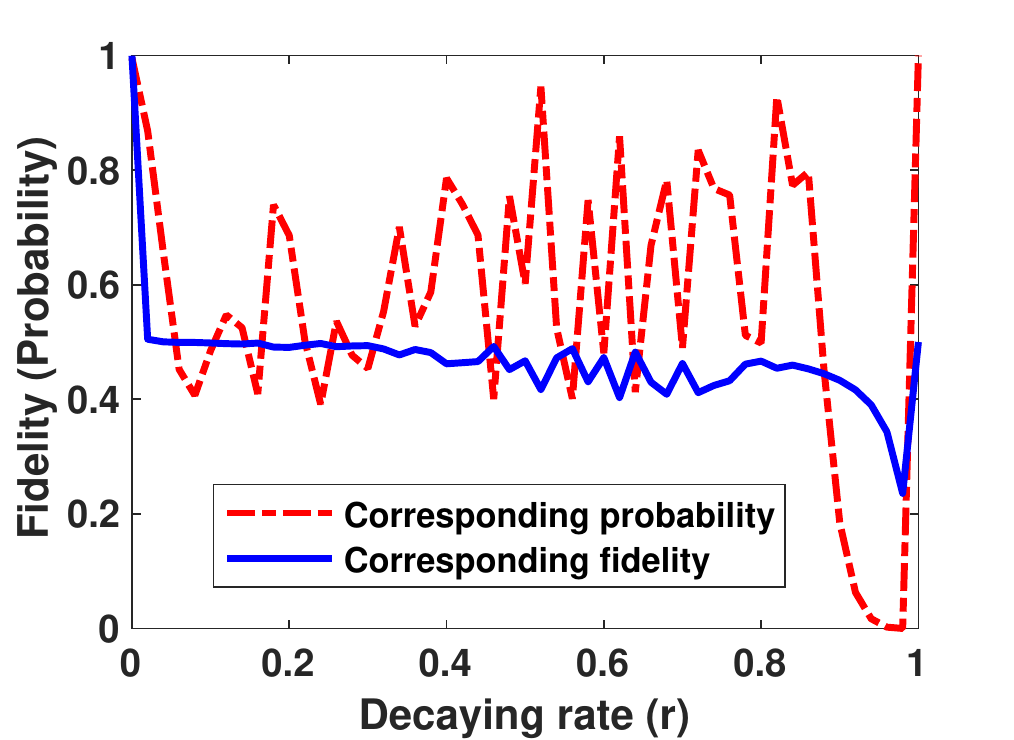}}
        \centering{(c)}
    \end{minipage}
\caption{(a) Average total QFI as a function of damping probability. (b) Optimum measurement and rotation angle to gain maximum QFI. (c) Corresponding fidelity and probability.}\label{fig2}
\end{figure*}

In this subsection, we study the protection of QFI of phase $\phi _{0} $ of GHZ state by fixing  $\gamma ={\pi }/{2}\ $in Eq. \eqref{eq5}. In Fig. \ref{fig2}, the number of qubits is fixed as $N=10$, the amount of damping probability $r$ is changing from 0 to 1 and the average total QFI is calculated according to Eq. \eqref{eq16}. For each amount of damping probability, we find the optimum measurement angle $\theta $ and rotation angle $\eta $ to gain the maximum QFI. When we use the optimum parameters to gain the maximum performance, we call the scheme as maximum QFFCR (MQFFCR). The do nothing (DN) scheme is the case where the states pass through the noise channel without any protection. The corresponding Fidelity and probability of maximum QFI are also given in Fig. \ref{fig2}(c). Also, for comparison, the maximum QFI of WMPPF is given as MWMPPF with the same parameters, where the pre weak measurement in WMPPF scheme is optimized to get the optimized QFI. 

As Fig. \ref{fig2}(a) depicted, the MQFFCR scheme protects QFI of phase completely and has great improvement over DN case even for intense decaying rates. After decaying rate $r=0.82$ the amount of QFI decreases and becomes less than 15 for $r=0.9$. It is easy to find from the Fig. \ref{fig2}(a) that the MQFFCR scheme is all higher than the MWMPPF scheme, which means by using the MQFFCR we can highly improve the measurement accuracy of the phase.

Fig. \ref{fig2}(b) is the optimum amounts of measurement angle $\theta $ and rotation angle $\eta $ to obtain maximum QFI. The amount of measurement angle is $0<\theta <{\pi \mathord{\left/ {\vphantom {\pi  2}} \right. \kern-\nulldelimiterspace} 2} $  for most amounts of decaying rate, which means to gain maximum QFI the WM scheme of measurement operators are applied. In addition, the corresponding fidelity and probability of maximum QFI scheme is given in Fig. \ref{fig2}(c). As one can see from Fig. \ref{fig2}(c), the fidelity of maximum QFI is 0.5 for most decaying rates and the amount of probability is changing between 0.5 to 0.95 for decaying rates $r\le 0.9$. Hence, protection of QFI, decrease the protection of the state to 50\%, and one cannot protect both QFI and fidelity completely at the same time. 

\subsubsection{ Maximum Fidelity}

\begin{figure*}[h!]
\begin{center}
   
    \begin{minipage}[t]{0.45\textwidth}
        \centering{\includegraphics[width=\textwidth]{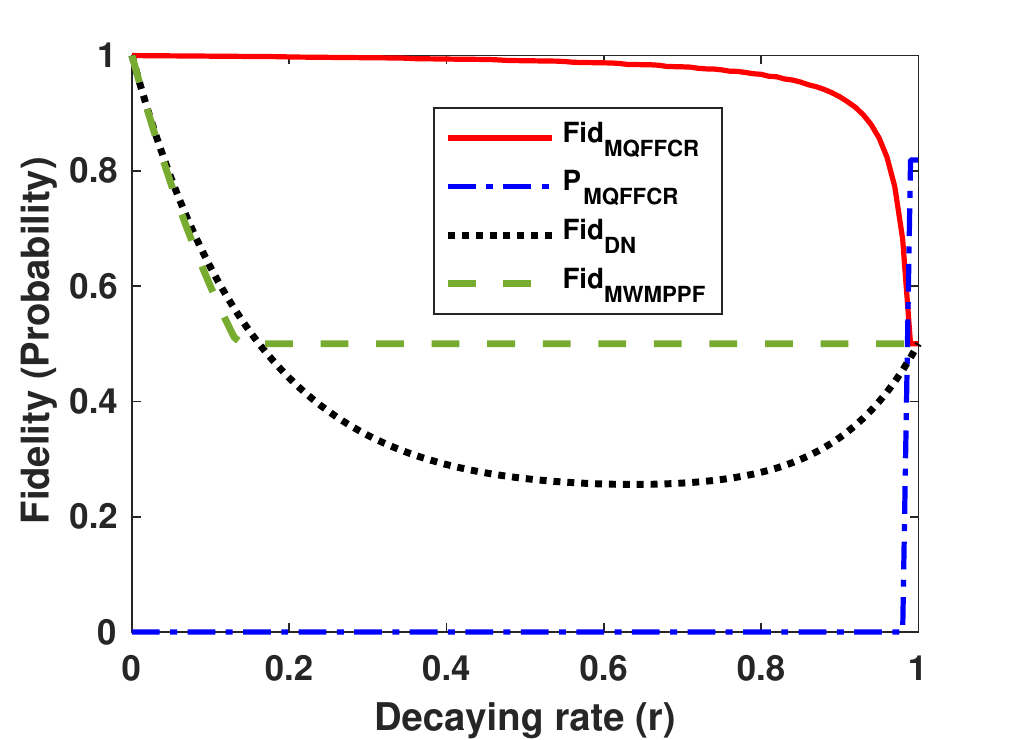}}
        \centering{(a)}
    \end{minipage}
    \hfill
    \begin{minipage}[t]{0.45\textwidth}
        \centering{\includegraphics[width=\textwidth]{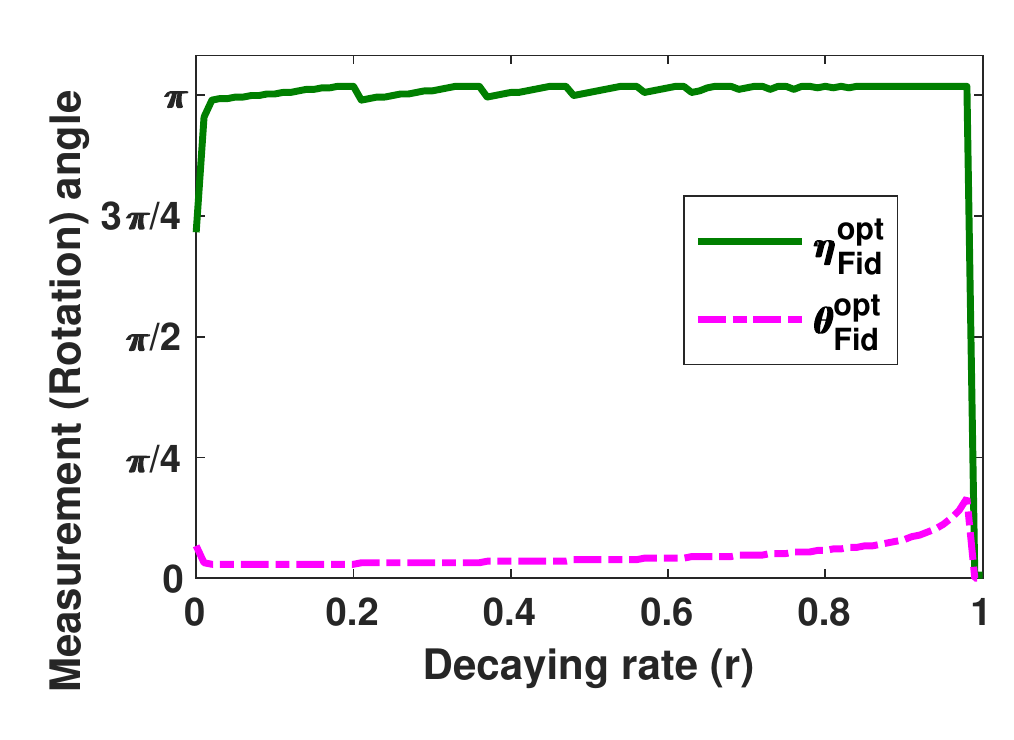}}
        \centering{(b)}
    \end{minipage}
    \end{center}
\caption{(a) Average total fidelity as a function of damping probability. (b) Optimum measurement and rotation angle to gain maximum QFI.}\label{fig3}
\end{figure*}

Here we study the behavior of maximum fidelity as a function of decaying rate $r$. The situation is same as Fig. \ref{fig2}, and the average total fidelity is calculated by Eq. \eqref{eq12}. Again, for each amount of decaying rate the optimum measurement and rotation angle is used to gain maximum fidelity; hence, we call the scheme as MQFFCR. Fig. \ref{fig3} shows the maximum fidelity and corresponding probability over 0 to 1 decaying rate. The corresponding measurement and rotation angles are also given.  

As Fig. \ref{fig3} shows, the maximized fidelity of MQFFCR is more than 98\% for decaying rates $r\le 0.7$. For higher decaying rates, the maximized fidelity decreases until it become 0.5 at $r=1$. As we said before, there is a trade-off between fidelity and probability as shown in Eq. \eqref{eq12} , which means by increasing one, the other will decrease. That is why the corresponding probability of maximum fidelity is zero for most decaying rates. In addition, Fig. \ref{fig3}(a) shows that MQFFCR can give much better protection compared to MWMPPF scheme for all decaying rates. Fig. \ref{fig3} (b) is the optimum rotation and measurement angle to gain maximum fidelity. As Fig. \ref{fig3} (b) depicted, to gain maximum fidelity the measurement angle must be small which means strong measurement or nearly projective measurement is needed. Also, the amount of rotation angle is near $\pi $ for most decaying rates.

\subsubsection{ Maximum probability}

In this subsection, we give the optimum condition to gain maximum probability. The behavior of maximum probability and corresponding fidelity as a function of decaying rate $r$ is given. The situation is same as Fig. \ref{fig2}, the average total probability is calculated as Eq. \eqref{eq9} and the average total fidelity as Eq. \eqref{eq12}. Again, for each damping probability the optimum measurement and rotation angle is used to gain maximum probability; hence, we call the scheme as MQFFCR. 

\begin{figure*}[h!]
\begin{center}
   
    \begin{minipage}[t]{0.45\textwidth}
        \centering{\includegraphics[width=\textwidth]{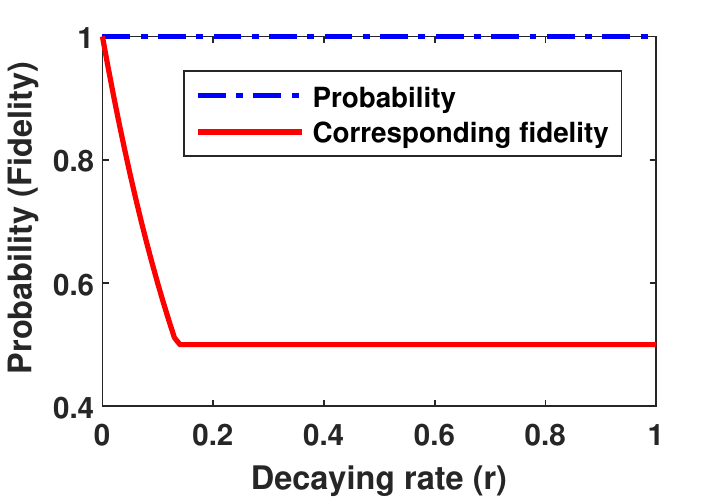}}
        \centering{(a)}
    \end{minipage}
    \hfill
    \begin{minipage}[t]{0.45\textwidth}
        \centering{\includegraphics[width=\textwidth]{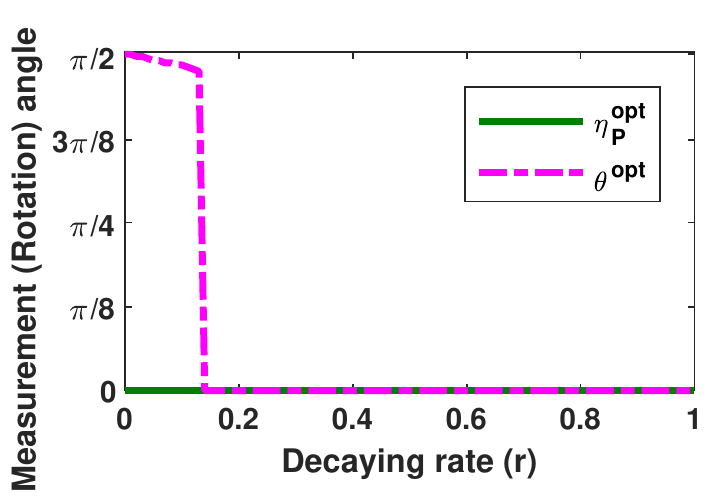}}
        \centering{(b)}
    \end{minipage}
    \end{center}
\caption{(a) Maximized probability as a function of decaying rate. (b) Optimum measurement and rotation angle to gain maximum probability.}\label{fig4}
\end{figure*}

In order to gain maximum probability $P=1$, the trace of the final density matrix must be $Tr(\rho _{f} )=1$. Hence, we derive the optimum rotation angle to gain maximum probability as:
\begin{equation} \label{eq17} 
\eta _{P}^{opt} \left(r,\theta \right)=-i\log \left(\frac{1+\sqrt{1-4\left(\cos ^{2} (\theta /2)+r\sin ^{2} (\theta /2)\right)\left((1-r)*\sin ^{2} (\theta /2)\right)} }{2\left(\cos ^{2} (\theta /2)+r\sin ^{2} (\theta /2)\right)} \right) 
\end{equation} 
    We note that, by setting rotation angle as Eq. \eqref{eq17}, the amount of probability $P$ is always 1, for all amounts of decaying rate $r$ and measurement angle $\theta $. Therefore, for each $r$ and fixed optimum rotation angle $\eta $, we find the maximum fidelity by changing $\theta $ from 0 to $\pi $ and plot the maximum probability and corresponding maximum fidelity as a function of decaying rate $r$ in Fig. \ref{fig4}. 

As Fig. \ref{fig4}(a) shows, probability 1 is achievable for all decaying rates. Although, the corresponding fidelity of maximum probability is 0.5 and only for low decaying rates $r\le 0.15$, we can gain fidelity more than 0.5.  As Fig. \ref{fig4}(b) demonstrates, to gain maximum probability for all decaying rates, rotation angle is always equal to 0. In other words, no rotation must apply if one wants to achieve the probability 1. Moreover, the optimum measurement angle for small decaying rates $r\le 0.15$ is near $\frac{\pi }{2} $ , which makes the measurement becomes weak measurement. While for $r>0.15$ the optimum measurement angle is 0, which makes the measurement becomes projective measurement.

\subsubsection{ Fidelity and probability comparison}

\begin{figure*}[h!]
\begin{center}
   
    \begin{minipage}[t]{0.45\textwidth}
        \centering{\includegraphics[width=\textwidth]{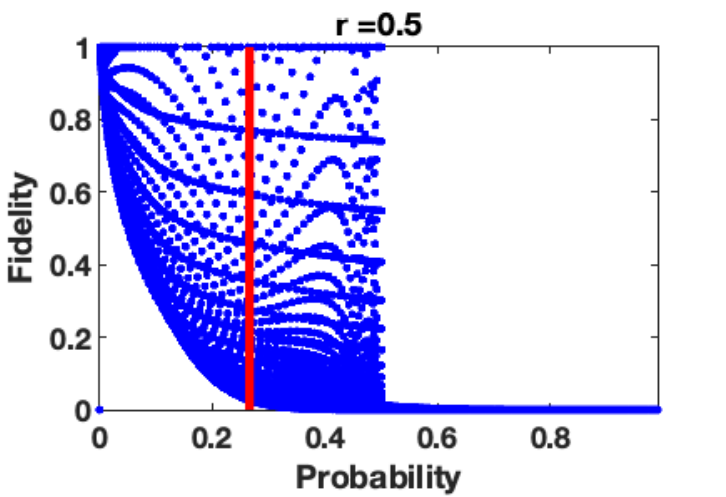}}
        \centering{(a)}
    \end{minipage}
    \hfill
    \begin{minipage}[t]{0.45\textwidth}
        \centering{\includegraphics[width=\textwidth]{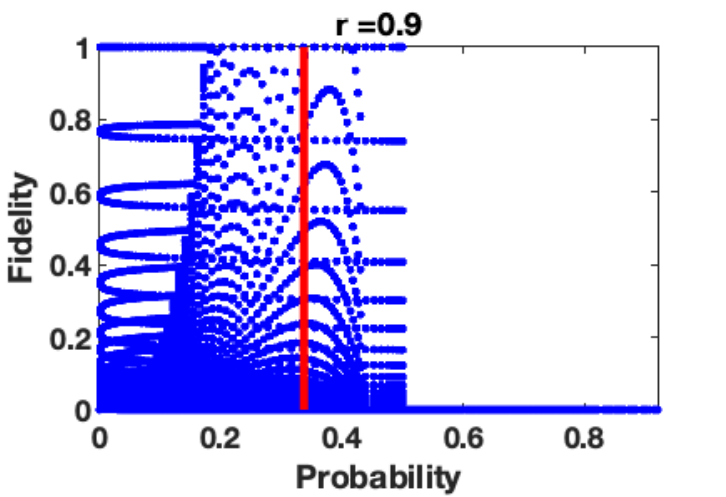}}
        \centering{(b)}
    \end{minipage}
    \end{center}
\caption{The relation between average total fidelity and average total probability. The solid red line represents the fidelity without any control. (a)  r= 0.5. (b)  r= 0.9.}\label{fig5}
\end{figure*}

As we explained before there is a trade-off between probability and fidelity of our scheme. In this subsection to show the relation between fidelity and probability, we change the amount of measurement and rotation angle from 0 to $\pi $ and plot corresponding average total fidelity Eq.\eqref{eq12} and average total probability Eq. \eqref{eq9} in Fig. \ref{fig5} for fixed amounts of decaying rate $r=\, 0.5$, and 0.9. The blue dots demonstrate the performance of the proposed QFFCR scheme for all independent amounts of measurement angle $\theta $ and rotation angle $\eta $. 

As Fig. \ref{fig5} depicted, the proposed scheme improves the amount of fidelity even for high decaying rates $r=\, 0.9$. However, for fidelity more than 0.5 the probability decreases to zero. Therefore, we can conclude that by setting optimum parameters, the maximum fidelity 0.5 is attainable with probability 1. One can gain fidelity more than 0.5 to 1 by giving up the probability (the probability is close to zero for fidelities more than 0.5).

\subsection{   Protection of generalized GHZ states}

\begin{figure*}[ht!]
    \centering
    \begin{minipage}[t]{0.36\textwidth}
        \centering{\includegraphics[width=\textwidth]{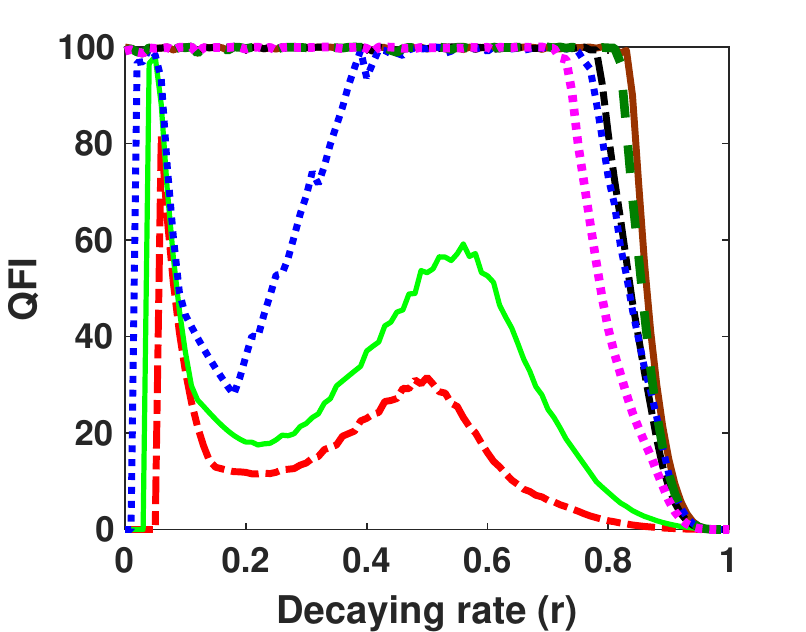}}
       \centering{(a)}
    \end{minipage}
    \hfill
    \begin{minipage}[t]{0.36\textwidth}
        \centering{\includegraphics[width=\textwidth]{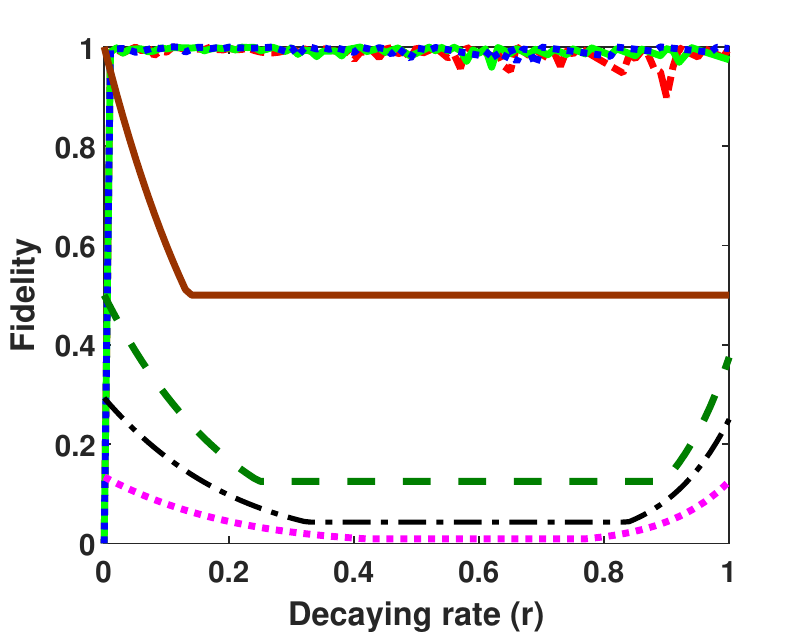}}
      	\centering{(b)}
    \end{minipage}
    \begin{minipage}[t]{0.26\textwidth}
        \centering{\includegraphics[width=1.2\textwidth]{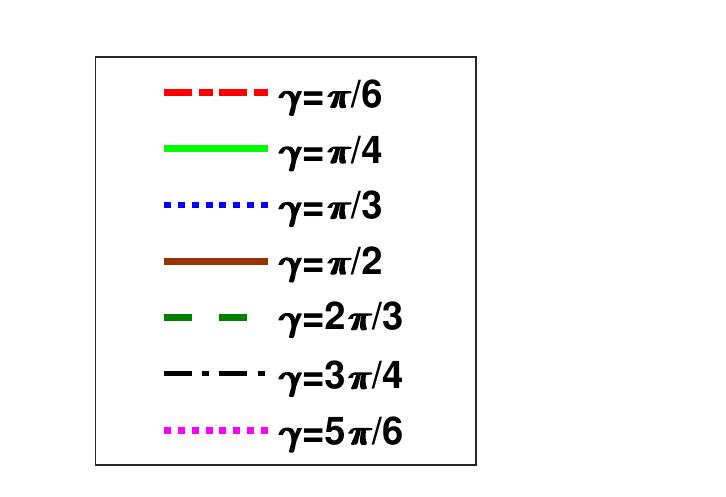}}
    \end{minipage}
\caption{Average total QFI of phase and fidelity for different generalize GHZ states against amplitude damping for $N$=10.}\label{fig6}
\end{figure*}

   Above we mainly discuss the \textit{N}-qubit GHZ state where $\left|\alpha \right|=\left|\beta \right|,\, \left(\gamma ={\pi \mathord{\left/ {\vphantom {\pi  2}} \right. \kern-\nulldelimiterspace} 2} \right)$ and $\phi _{0} =0$ in Eq. \eqref{eq5}. If  $\left|\alpha \right|\ne \left|\beta \right|,\, \left(\gamma \ne {\pi \mathord{\left/ {\vphantom {\pi  2}} \right. \kern-\nulldelimiterspace} 2} \right)$, it is non-maximally entangled state or generalized GHZ state. To find the behavior of proposed protection control for different initial generalized GHZ states, Fig. \ref{fig6} shows the average total QFI for different amount of $\gamma $.

As Fig. \ref{fig6} demonstrated, when $\alpha <\beta \, \, \left({\pi \mathord{\left/ {\vphantom {\pi  2\le \gamma }} \right. \kern-\nulldelimiterspace} 2\le \gamma } <\pi \right)$, the QFFCR can protect the QFI of phase for generalized GHZ states effectively for $r\le 0.8$. When $\gamma $ is close to ${\pi \mathord{\left/ {\vphantom {\pi  2}} \right. \kern-\nulldelimiterspace} 2} $, the complete protection of QFI lasts for higher decaying rates, until $r=0.9$. Also, the smaller the angle $\gamma $ is, the lower QFI achieved.   

As we show in subsection 3.1.2, by maximizing fidelity, the amount of probability decreases to zero. Hence, we study the behavior of fidelity for 10 qubits generalized GHZ state in case of maximizing probability. By setting the rotation angle as Eq. \eqref{eq17} , we gain the probability 1, and maximum fidelity by finding optimum measurement angle $\theta $ for each decaying rate $r$. The Maximized fidelity according to probability 1 for different generalized GHZ states is given in Fig. \ref{fig6}(b).  As Fig. \ref{fig6}(b) depicted, when $\alpha >\beta \, \, \left(0<\gamma \le {\pi \mathord{\left/ {\vphantom {\pi  2}} \right. \kern-\nulldelimiterspace} 2} \right)$, the QFFCR can protect the generalized GHZ states completely and gain the average fidelity more than 99\% and probability 1, even for intense decaying rates. Here again we note that, the corresponding probability of the plotted fidelities in Fig. \ref{fig6}(b) is 1.

Therefore, we can conclude that in case of generalized GHZ states protection, one cannot protect both the state and fidelity at the same time. However, by protecting each, the QFFCR give the best results for special states. The states that $\alpha >\beta \, \, \left(0<\gamma \le {\pi \mathord{\left/ {\vphantom {\pi  2}} \right. \kern-\nulldelimiterspace} 2} \right)$ can be protected completely with almost complete fidelity and probability 1, even for high decaying rates; and the QFI of the states $\alpha <\beta \, \, \left({\pi \mathord{\left/ {\vphantom {\pi  2\le \gamma }} \right. \kern-\nulldelimiterspace} 2\le \gamma } <\pi \right)$, is protected significantly for $r\le 0.8$.

\section{ Conclusion}

In this paper, we study the protection of \textit{N}-qubit GHZ and generalized GHZ states in ADC by means of weak measurement and offsetting operations. Before the state enters the ADC, we apply a weak measurement to obtain information about the state, and offsetting operation according to the result of the weak measurement to make the state less vulnerable to the effects of ADC. After the ADC, the same offsetting operation and a rotation operation are applied to retrieve the state. The final expressions to calculate the average total probability, fidelity and QFI for both GHZ state and generalized GHZ states are given. According to formulas we have shown that the proposed scheme can effectively protect the average QFI and the state of the N-qubit GHZ state and some generalized GHZ states. In case of protecting GHZ states, there is a trade-off between probability and fidelity, and with probability 1 the maximum fidelity 0.5 is attainable. For generalized GHZ states, when $\alpha <\beta \, \, \left({\pi \mathord{\left/ {\vphantom {\pi  2\le \gamma }} \right. \kern-\nulldelimiterspace} 2\le \gamma } <\pi \right)$ , the proposed scheme can protect the QFI of phase effectively for  $r\le 0.8$; and when $\alpha >\beta \, \, \left(0<\gamma \le {\pi \mathord{\left/ {\vphantom {\pi  2}} \right. \kern-\nulldelimiterspace} 2} \right)$ the states can be protected completely with almost complete fidelity and probability 1, even for high decaying rates. Comparing our scheme with WMPPF shows the significant improvement of QFI protection in a probabilistic manner. Although WMPPF has the probability 1, our scheme notably increase the amount of QFI with acceptable probability even for intense decaying rates. Furthermore, our scheme is entirely feasible with current technology. The experimental realization of the weak measurement and correction rotation used in our scheme is discussed in \cite{lab19}.

\section*{Appendix}

\section*{Density matrix of N-qubit GHZ state at each step of QFFCR }

Here we derive the density matrix of the system at each step of the control procedure.\textbf{}

Here we assume that $i$ qubits are measured by $M_{0}^{i} $ and $j$ qubits are measured by $M_{1}^{j} $, where $M_{0}^{i} =M_{0} $ and $M_{1}^{j} =M_{1} $ are given in Eq. \eqref{eq1}. Hence, we define $i\in {\rm A} ,\, \, j\in {\rm B} ,\, \, {\rm A} \cap {\rm B} =\emptyset ,\, \, {\rm A} \cup {\rm B} =\left\{1,\cdots ,N\right\}\, $, where ${\rm A} $ and $B$indicate a concrete combination of the qubits measured by $M_{0}^{i} $ and $M_{1}^{j} $. ${\rm A} $ has $k$ elements with $k=\left\{0,1,\ldots ,N\right\}$ and $B$ has $N-k$ elements. One case of measurement operator can be $M_{S_{k} } =\left[\left(M_{0}^{i} \right)^{\otimes k} \otimes \left(M_{1}^{j} \right)^{\otimes N-k} \right]_{S_{k} } $. If we fix the amount of $k$, $M_{S_{k} } $has ${\rm C}_{N}^{k} =\frac{N!}{k!(N-k)!} $   combinations by changing the qubits, while the fixed number of qubits measured by $M_{0}^{i} $ is $k$ and the fixed number of qubits measured by $M_{1}^{j} $ is $N-k$. The total combinations with all possible $k$is ${\rm C}_{N}^{0} +{\rm C}_{N}^{1} +\cdots +{\rm C}_{N}^{N} =2^{N} $. The weak measurement set is a complete measurement, which $I^{\otimes 2N} =\sum _{k=0}^{N}\sum _{S_{k} =1}^{{\rm C}_{N}^{k} }M_{S_{k} }^{\dag } M_{S_{k} }   $. By applying the weak measurement set for $N$ qubits, there are $2^{N} $cases which corresponds to a combination of $\left(M_{0}^{i} \right)^{\otimes k} \otimes \left(M_{1}^{j} \right)^{\otimes N-k} $, where $k$ qubits are measured by $M_{0} $ and $N-k$ qubits measured by $M_{1} $ for $k=0,1,\cdots ,N$. Hence the weak measurement operator can be written as
\begin{equation} \label{18)} 
\begin{array}{l} {M_{S_{k} } =\left[\left(\cos (\theta /2){\left| 0 \right\rangle} {\left\langle 0 \right|} +\sin (\theta /2){\left| 1 \right\rangle} {\left\langle 1 \right|} \right)^{\otimes k} \otimes \left(\sin (\theta /2){\left| 0 \right\rangle} {\left\langle 0 \right|} +\cos (\theta /2){\left| 1 \right\rangle} {\left\langle 1 \right|} \right)^{\otimes N-k} \right]_{S_{k} } } \\ {=\left[\left(\begin{array}{cc} {\cos (\theta /2)} & {0} \\ {0} & {\sin (\theta /2)} \end{array}\right)^{\otimes k} \otimes \left(\begin{array}{cc} {\sin (\theta /2)} & {0} \\ {0} & {\cos (\theta /2)} \end{array}\right)^{\otimes N-k} \right]_{S_{k} } } \end{array} 
\end{equation} 
In step 2 of the control procedure, the offsetting operations $F_{0}^{i} =F_{0} $ and $F_{1}^{j} =F_{1} $ are applied to qubits according to the result of the weak measurements $M_{0}^{i} $ and $M_{1}^{j} $ , respectively. Hence, the weak measurement and offsetting coupled operation is
\begin{equation} \label{19)} 
\begin{array}{l} {U_{M-F} =\left[\left(F_{0}^{i} \right)^{\otimes k} \otimes \left(F_{1}^{j} \right)^{\otimes N-k} \left(M_{0}^{i} \right)^{\otimes k} \otimes \left(M_{1}^{j} \right)^{\otimes N-k} \right]_{S_{k} } } \\ {=\left[\left(\begin{array}{cc} {\cos (\theta /2)} & {0} \\ {0} & {\sin (\theta /2)} \end{array}\right)^{\otimes k} \otimes \left(\begin{array}{cc} {0} & {\cos (\theta /2)} \\ {\sin (\theta /2)} & {0} \end{array}\right)^{\otimes N-k} \right]_{S_{k} } } \end{array} 
\end{equation} 
The density matrix of the system after weak measurement and offsetting operations will become
\begin{equation} \label{ZEqnNum218329} 
\left(\begin{array}{l} {\begin{array}{ccc} {\ddots } & {} & {} \\ {} & {W} & {} \\ {} & {} & {\ddots } \end{array}\, \, \, \, \, \begin{array}{ccc} {} & {} & {{\mathinner{\mkern2mu\raise1pt\hbox{.}\mkern2mu\raise4pt\hbox{.}\mkern2mu\raise7pt\hbox{.}\mkern1mu}} } \\ {} & {X} & {} \\ {{\mathinner{\mkern2mu\raise1pt\hbox{.}\mkern2mu\raise4pt\hbox{.}\mkern2mu\raise7pt\hbox{.}\mkern1mu}} } & {} & {} \end{array}} \\ {\begin{array}{ccc} {} & {} & {{\mathinner{\mkern2mu\raise1pt\hbox{.}\mkern2mu\raise4pt\hbox{.}\mkern2mu\raise7pt\hbox{.}\mkern1mu}} } \\ {} & {Y} & {} \\ {{\mathinner{\mkern2mu\raise1pt\hbox{.}\mkern2mu\raise4pt\hbox{.}\mkern2mu\raise7pt\hbox{.}\mkern1mu}} } & {} & {} \end{array}\, \, \, \, \, \, \begin{array}{ccc} {\ddots } & {} & {} \\ {} & {Z} & {} \\ {} & {} & {\ddots } \end{array}} \end{array}\right) 
\end{equation} 
where only four elements $W,\, X,\, Y,\, Z$ are non-zero and all the other elements of the density matrix are zero. The amounts of four non-zero elements of the density matrix are 
\begin{equation} \label{21)} 
\begin{array}{l} {W=\left|\alpha \right|^{2} \left[\left(\begin{array}{cc} {\cos ^{2} (\theta /2)} & {0} \\ {0} & {0} \end{array}\right)^{\otimes k} \otimes \left(\begin{array}{cc} {0} & {0} \\ {0} & {\sin ^{2} (\theta /2)} \end{array}\right)^{\otimes N-k} \right]_{S_{k} } ,} \\ {Z=\left|\beta \right|^{2} \left[\left(\begin{array}{cc} {0} & {0} \\ {0} & {\sin ^{2} (\theta /2)} \end{array}\right)^{\otimes k} \otimes \left(\begin{array}{cc} {\cos ^{2} (\theta /2)} & {0} \\ {0} & {0} \end{array}\right)^{\otimes N-k} \right]_{S_{k} } ,} \\ {X=\alpha \beta ^{*} \left[\left(\begin{array}{cc} {0} & {0} \\ {\cos (\theta /2)\sin (\theta /2)} & {0} \end{array}\right)^{\otimes k} \otimes \left(\begin{array}{cc} {0} & {\cos (\theta /2)\sin (\theta /2)} \\ {0} & {0} \end{array}\right)^{\otimes N-k} \right]_{S_{k} } ,} \\ {Y=\alpha ^{*} \beta \left[\left(\begin{array}{cc} {0} & {\cos (\theta /2)\sin (\theta /2)} \\ {0} & {0} \end{array}\right)^{\otimes k} \otimes \left(\begin{array}{cc} {0} & {0} \\ {\cos (\theta /2)\sin (\theta /2)} & {0} \end{array}\right)^{\otimes N-k} \right]_{S_{k} } } \end{array} 
\end{equation} 
Now the state goes through ADC. In this paper we assume that $N$ qubits go through $N$independent amplitude damping channels which have the same damping probability $r_{1} =r_{2} =\ldots =r_{N} =r$. For a general single-qubit $\rho $, amplitude damping can also be written as
\begin{equation} \label{ZEqnNum617364} 
\rho \to \varepsilon _{AD} \left(\rho \right)=e_{0} \, \rho \, e_{0}^{\dag } +e_{1} \, \rho \, e_{1}^{\dag }  
\end{equation} 
According to Eq. \eqref{eq3} and \eqref{ZEqnNum617364}, we conclude following effects of ADC 
\begin{equation} \label{23)} 
\begin{array}{l} \scriptstyle{\left|\left. 0\right\rangle \right. \left\langle \left. 0\right|\right. =\left(\begin{array}{cc} {1} & {0} \\ {0} & {0} \end{array}\right)\to \left(\begin{array}{cc} {1} & {0} \\ {0} & {0} \end{array}\right),\, \, \, \left|\left. 1\right\rangle \right. \left\langle \left. 1\right|\right. =\left(\begin{array}{cc} {1} & {0} \\ {0} & {0} \end{array}\right)\to \left(\begin{array}{cc} {r} & {0} \\ {0} & {1-r} \end{array}\right)} \\ \scriptstyle {\left|\left. 1\right\rangle \right. \left\langle \left. 0\right|\right. =\left(\begin{array}{cc} {0} & {0} \\ {1} & {0} \end{array}\right)\to \left(\begin{array}{cc} {0} & {0} \\ {\sqrt{1-r} } & {0} \end{array}\right),\, \, \left|\left. 0\right\rangle \right. \left\langle \left. 1\right|\right. =\left(\begin{array}{cc} {0} & {1} \\ {0} & {0} \end{array}\right)\to \left(\begin{array}{cc} {0} & {\sqrt{1-r} } \\ {0} & {0} \end{array}\right)} \end{array} 
\end{equation}

Hence, the elements of density matrix in Eq. \eqref{ZEqnNum218329} after passing through ADC become
\begin{equation} \label{24)} 
\begin{array}{l} {W_{e} =\left|\alpha \right|^{2} \left[\left(\begin{array}{cc} {\cos ^{2} (\theta /2)} & {0} \\ {0} & {0} \end{array}\right)^{\otimes k} \otimes \left(\begin{array}{cc} {\sin ^{2} (\theta /2).\, r} & {0} \\ {0} & {\sin ^{2} (\theta /2).(1-r)} \end{array}\right)^{\otimes N-k} \right]_{S_{k} } ,} \\ \\ \scriptstyle{X_{e} =\alpha \beta ^{*} \left[\left(\begin{array}{cc} {0} & {0} \\ {\cos (\theta /2)\sin (\theta /2)\sqrt{1-r} } & {0} \end{array}\right)^{\otimes k} \otimes \left(\begin{array}{cc} {0} & {\cos (\theta /2)\sin (\theta /2)\sqrt{1-r} } \\ {0} & {0} \end{array}\right)^{\otimes N-k} \right]_{S_{k} } ,} \\ \\ \scriptstyle {Y_{e} =\alpha ^{*} \beta \left[\left(\begin{array}{cc} {0} & {\cos (\theta /2)\sin (\theta /2)\sqrt{1-r} } \\ {0} & {0} \end{array}\right)^{\otimes k} \otimes \left(\begin{array}{cc} {0} & {0} \\ {\cos (\theta /2)\sin (\theta /2)\sqrt{1-r} } & {0} \end{array}\right)^{\otimes N-k} \right]_{S_{k} } ,} \\ \\ {Z_{e} =\left|\beta \right|^{2} \left[\left(\begin{array}{cc} {\sin ^{2} (\theta /2).r} & {0} \\ {0} & {\sin ^{2} (\theta /2).\, (1-r)} \end{array}\right)^{\otimes k} \otimes \left(\begin{array}{cc} {\cos ^{2} (\theta /2)} & {0} \\ {0} & {0} \end{array}\right)^{\otimes N-k} \right]_{S_{k} } } \end{array} 
\end{equation} 
After the ADC, the same offsetting operators are applied. So the state of the system after offsetting operations in step 4 of the control procedure are 
\begin{equation} \label{25)} 
\begin{array}{l} {W_{e-f} =\left|\alpha \right|^{2} \left[\left(\begin{array}{cc} {\cos ^{2} (\theta /2)\, } & {0} \\ {0} & {0} \end{array}\right)^{\otimes k} \otimes \left(\begin{array}{cc} {\sin ^{2} (\theta /2).(1-r)} & {0} \\ {0} & {\sin ^{2} (\theta /2).\, r} \end{array}\right)^{\otimes N-k} \right]_{S_{k} } ,} \\ \\ \scriptstyle{X_{e-f} =\alpha \beta ^{*} \left[\left(\begin{array}{cc} {0} & {\cos (\theta /2)\sin (\theta /2)\sqrt{1-r} \, } \\ {0} & {0} \end{array}\right)^{\otimes k} \otimes \left(\begin{array}{cc} {0} & {0} \\ {\cos (\theta /2)\sin (\theta /2)\sqrt{1-r} \, } & {0} \end{array}\right)^{\otimes N-k} \right]_{S_{k} } ,} \\ \\ \scriptstyle {Y_{e-f} =\alpha ^{*} \beta \left[\left(\begin{array}{cc} {0} & {0} \\ {\cos (\theta /2)\sin (\theta /2)\sqrt{1-r} \, } & {0} \end{array}\right)^{\otimes k} \otimes \left(\begin{array}{cc} {0} & {0} \\ {\cos (\theta /2)\sin (\theta /2)\sqrt{1-r} \, } & {0} \end{array}\right)^{\otimes N-k} \right]_{S_{k} } ,} \\ \\ {Z_{e-f} =\left|\beta \right|^{2} \left[\left(\begin{array}{cc} {\sin ^{2} (\theta /2).r} & {0} \\ {0} & {\sin ^{2} (\theta /2).\, (1-r)} \end{array}\right)^{\otimes k} \otimes \left(\begin{array}{cc} {0} & {0} \\ {0} & {\cos ^{2} (\theta /2)} \end{array}\right)^{\otimes N-k} \right]_{S_{k} } } \end{array} 
\end{equation} 
At the final step, we apply the correction rotation to retrieve the state. After the whole process of state protection, we can normalize the final state as:
\begin{equation} \label{ZEqnNum643971} 
\rho _{f} =\frac{1}{P} \tilde{\rho }=\frac{1}{P} \left(\begin{array}{ccc} {A} & {0} & {D} \\ {0} & {E} & {0} \\ {C} & {0} & {B} \end{array}\right) 
\end{equation} 
where$\tilde{\rho }$ is the unnormalized final state and $P$ is the probability of appearing $\rho _{f} $.  In $\rho _{f} $ the four elements $A,\, B,\, C,\, D$ are four corners of the density matrix with bases $\left|\left. 0\right\rangle \right. \left\langle \left. 0\right|\right. ^{\otimes N} ,\, \left|\left. 1\right\rangle \right. \left\langle \left. 1\right|\right. ^{\otimes N} ,\, \left|\left. 1\right\rangle \right. \left\langle \left. 0\right|\right. ^{\otimes N} ,\, \left|\left. 0\right\rangle \right. \left\langle \left. 1\right|\right. ^{\otimes N} $  respectively. $E$ is a diagonal matrix with $2^{N} -2$ elements.

Here again we assume that $k$ qubits are measured by $M_{0} $ and $N-k$ qubits are measured by $M_{1} $. We give the elements of the final state in three cases: 

\begin{enumerate}
\item  When $k=0$, all the qubits are measured by $M_{1} $, the elements of the final state are
\begin{equation} \label{27)} 
\begin{array}{l} {A_{0} =\left[\left|\alpha \right|^{2} \sin ^{2N} \left(\frac{\theta }{2} \right)e^{-Ni\eta } \left(1-r\right)^{N} \right]{\left| 0 \right\rangle} {\left\langle 0 \right|} ^{\otimes N} } \\ {B_{0} =\left[e^{Ni\eta } \left(\left|\alpha \right|^{2} r^{N} \sin ^{2N} \left(\frac{\theta }{2} \right)+\left|\beta \right|^{2} \cos ^{2N} \left(\frac{\theta }{2} \right)\right)\right]{\left| 1 \right\rangle} {\left\langle 1 \right|} ^{\otimes N} } \\ {C_{0} =D_{0}^{\dag } =\alpha ^{*} \beta \left(1-r\right)^{{\raise0.7ex\hbox{$ N $}\!\mathord{\left/ {\vphantom {N 2}} \right. \kern-\nulldelimiterspace}\!\lower0.7ex\hbox{$ 2 $}} } \frac{\sin ^{N} \left(\theta \right)}{2^{N} } } \\ {E_{0} =\alpha ^{2} \sin ^{2N} \left(\frac{\theta }{2} \right)*}\\{\left[\left(e^{-i\eta } (1-r){\left| 0 \right\rangle} {\left\langle 0 \right|} +e^{i\eta } r{\left| 1 \right\rangle} {\left\langle 1 \right|} \right)^{\otimes N} -e^{-Ni\eta } (1-r)^{N} {\left| 0 \right\rangle} {\left\langle 0 \right|} ^{\otimes N} -e^{Ni\eta } r{\left| 1 \right\rangle} {\left\langle 1 \right|} ^{\otimes N} \right]}\end{array} 
\end{equation} 
\end{enumerate}
The probability of given final state is 
\begin{equation} \label{28)} 
P_{0} =\alpha ^{2} \sin ^{2N} \left(\frac{\theta }{2} \right)\left(re^{Ni\eta } +\left(1-r\right)e^{-Ni\eta } \right)^{N} +\beta ^{2} \cos ^{2N} \left(\frac{\theta }{2} \right)e^{Ni\eta }  
\end{equation} 

\begin{enumerate}
\item  If we assume $k=N$, which means all the qubits are measured by $M_{0} $. In this case the elements of the final state are:
\begin{equation} \label{29)} 
\begin{array}{l} {A_{N} =\left[e^{Ni\eta } \left(\alpha ^{2} \cos ^{2N} \left(\frac{\theta }{2} \right)+\beta ^{2} r^{N} \sin ^{2N} \left(\frac{\theta }{2} \right)\right)\right]{\left| 0 \right\rangle} {\left\langle 0 \right|} ^{\otimes N} } \\ {B_{N} =\left[\beta ^{2} \sin ^{2N} \left(\frac{\theta }{2} \right)e^{-Ni\eta } \left(1-r\right)^{N} \right]{\left| 1 \right\rangle} {\left\langle 1 \right|} ^{\otimes N} } \\ {C_{N} =D_{N}^{\dag } =\alpha ^{*} \beta \left(1-r\right)^{{\raise0.7ex\hbox{$ N $}\!\mathord{\left/ {\vphantom {N 2}} \right. \kern-\nulldelimiterspace}\!\lower0.7ex\hbox{$ 2 $}} } \frac{\sin ^{2N} \left(\theta \right)}{2^{N} } } \\ {E_{N} =\beta ^{2} \sin ^{2N} \left(\frac{\theta }{2} \right)*}\\{\left[\left(e^{i\eta } r{\left| 0 \right\rangle} {\left\langle 0 \right|} +e^{-i\eta } (1-r){\left| 1 \right\rangle} {\left\langle 1 \right|} \right)^{\otimes N} -e^{Ni\eta } r{\left| 0 \right\rangle} {\left\langle 0 \right|} ^{\otimes N} -e^{-Ni\eta } (1-r)^{N} {\left| 1 \right\rangle} {\left\langle 1 \right|} ^{\otimes N} \right]} \end{array} 
\end{equation} 
\end{enumerate}
With probability 
\begin{equation} \label{30)} 
P_{N} =\alpha ^{2} \cos ^{2N} \left(\frac{\theta }{2} \right)e^{Ni\eta } +\beta ^{2} \sin ^{2N} \left(\frac{\theta }{2} \right)\left(re^{i\eta } +\left(1-r\right)e^{-i\eta } \right)^{N}  
\end{equation} 

\begin{enumerate}
\item  If we assume $k$ qubits are measured by $M_{0} $ and $N-k$ qubits are measured by $M_{1} $ for $k=1$ to $N-1$, the elements of the final state are:
\begin{equation} \label{31)} 
\begin{array}{l} {A_{S_{k} } =\left[\left|\alpha \right|^{2} \cos ^{2k} \left(\frac{\theta }{2} \right)\sin ^{2\left(N-k\right)} \left(\frac{\theta }{2} \right)\left(1-r\right)^{N-k} e^{\left(2k-N\right)i\eta } \right]{\left| 0 \right\rangle} {\left\langle 0 \right|} ^{\otimes N} } \\ {B_{S_{k} } =\left[\left|\beta \right|^{2} \sin ^{2k} \left(\frac{\theta }{2} \right)\cos ^{2\left(N-k\right)} \left(\frac{\theta }{2} \right)e^{\left(N-2k\right)i\eta } \left(1-r\right)^{k} \right]{\left| 1 \right\rangle} {\left\langle 1 \right|} ^{\otimes N} } \\ {C_{S_{k} } =D_{S_{k} }^{\dag } =\alpha ^{*} \beta \left(1-r\right)^{{\raise0.7ex\hbox{$ N $}\!\mathord{\left/ {\vphantom {N 2}} \right. \kern-\nulldelimiterspace}\!\lower0.7ex\hbox{$ 2 $}} } \frac{\sin ^{N} \left(\theta \right)}{2^{N} } } \\ {E_{S_{k} } =\left|\alpha \right|^{2} \sin ^{2\left(N-k\right)} \left(\frac{\theta }{2} \right)\cos ^{2k} \left(\frac{\theta }{2} \right)*}\\{\left[e^{ik\eta } {\left| 0 \right\rangle} {\left\langle 0 \right|} ^{\otimes k} \otimes \left(e^{-i\eta } (1-r){\left| 0 \right\rangle} {\left\langle 0 \right|} +e^{i\eta } r{\left| 1 \right\rangle} {\left\langle 1 \right|} \right)^{\otimes N-k} \right]+} \\{\left|\beta \right|^{2} \sin ^{2k} \left(\frac{\theta }{2} \right)\cos ^{2\left(N-k\right)} \left(\frac{\theta }{2} \right)*}\\{\left[\left(e^{i\eta } r{\left| 0 \right\rangle} {\left\langle 0 \right|} +e^{-i\eta } (1-r){\left| 1 \right\rangle} {\left\langle 1 \right|} \right)^{\otimes k} \otimes \left(e^{i\eta \left(N-k\right)} {\left| 1 \right\rangle} {\left\langle 1 \right|} ^{\otimes N-k} \right)\right]}  {-A_{S_{k} } -B_{S_{k} } } \end{array} 
\end{equation} 
\end{enumerate}
With probability:
\begin{equation} \label{32)} 
\begin{array}{l} {P_{S_{k} } =\alpha ^{2} \sin ^{2\left(N-k\right)} \left(\frac{\theta }{2} \right)\cos ^{2k} \left(\frac{\theta }{2} \right)e^{ki\eta } \left(re^{i\eta } +\left(1-r\right)e^{-i\eta } \right)^{N-k} } \\ {+\beta ^{2} \cos ^{2\left(N-k\right)} \left(\frac{\theta }{2} \right)e^{\left(N-k\right)i\eta } \sin ^{2k} \left(\frac{\theta }{2} \right)\left(re^{i\eta } +\left(1-r\right)e^{-i\eta } \right)^{N-k} } \end{array} 
\end{equation} 

\section*{Acknowledgments}

This work was partially supported by the National Natural Science Foundation of China under Grant  61973290 and Grant 61720106009. The research of J. J. Nieto has been partially supported by the Agencia Estatal de Investigacion (AEI) of Spain, project PID2020-113275GB-I00, co-financed by the European Fund for Regional Development (FEDER); and by Xunta de Galicia under grant ED431C 2019/02.

\section*{Acknowledgments}

\end{document}